\documentclass[runningheads,a4paper]{llncs}

\usepackage{amssymb}
\usepackage{graphicx}
\usepackage{here}

\usepackage{url}
\urldef{\mailsa}\path|{paul.feuto_njonko, sylviane.cardey, peter.greenfield}@univ-fcomte.fr|
\urldef{\mailsb}\path|walid.elabed@globaldataexcellence.com|
\newcommand{\keywords}[1]{\par\addvspace\baselineskip
\noindent\keywordname\enspace\ignorespaces#1}

\begin{document}

\mainmatter  

\title{RuleCNL: A Controlled Natural Language for Business Rule Specifications}

\titlerunning{ RuleCNL: A Controlled Natural Language for Business Rule Specifications}

\author{Paul Brillant Feuto Njonko\inst{1}
\and Sylviane Cardey\inst{1}
\and Peter Greenfield\inst{1}
\and Walid El Abed\inst{2}}

\authorrunning{Paul Brillant Feuto Njonko et al.}

\institute{Centre Tesni\`{e}re - \'{E}quipe d'Accueil  2283 \\
Universit\'{e} de Franche-Comt\'{e} - UFR SLHS\\
30, rue M\'{e}gevand, 25030 Besan\c{c}on Cedex, France\\
\mailsa\\
\and Global Data Excellence Ltd.\\
Geneva, Switzerland\\
\mailsb}

%
%

\toctitle{Lecture Notes in Computer Science}
\tocauthor{Authors' Instructions}
\maketitle

\begin{abstract}
Business rules represent the primary means by which companies define their business, perform their actions in order to reach their objectives. Thus, they need to be expressed unambiguously to avoid inconsistencies between business stakeholders and formally in order to be machine-processed.  A promising solution is the use of a controlled natural language (CNL) which is a good mediator between natural and formal languages. This paper presents RuleCNL, which is a CNL for defining business rules. Its core feature is the alignment of the business rule definition with the business vocabulary which ensures  traceability and  consistency with the business domain. The RuleCNL tool provides  editors that assist end-users in the writing process and automatic mappings into the Semantics of Business Vocabulary and Business Rules (SBVR) standard. SBVR is grounded in first order logic and includes constructs called semantic formulations that structure the meaning of rules. 
\end{abstract}

\section{Introduction}

Nowadays, companies are facing much pressure due to  competition and growth, which requires frequent adaptation of their business rules. However, for a couple of decades, business rules have been hard-coded in automated business processes, information systems and often inconsistently so. Thus, changing or modifying business rules inevitably requires software engineers'  intervention because they are inaccessible to business experts (e.g. healthcare experts, finance experts, etc.) who understand the actual problem domain and are responsible for finding solutions. As a result, companies cannot keep pace with the changing business environment.

The business rule approach (BRA) has evolved over the years in order to solve the deficiency described above \cite{Ronald2003} \cite{Babara}. It claims that all business rules should be collected and explicitly represented in a centralized application called business rule management system (BRMS). Many formal business rule languages  have been devised allowing companies to  define their business rules explicitly and formally. However, since such rules have to be created and/or verified by domain experts who are mostly not familiar with formal notations, a promising solution is the use of a controlled natural language (CNL) that can serve as a front-end interface and provide automatic mappings into formal notations. Thus, this paper presents RuleCNL for expressing business rules. Its core feature is the alignment of the business rule definition with the business vocabulary which ensures  traceability and consistency with the business domain. The underlying natural language (NL) in this paper is English but RuleCNL also works with French.  The RuleCNL tool provides  editors that assist end-users in the writing process and provides automatic mappings into the Semantics of Business Vocabulary and Business Rules (SBVR) standard. SBVR is grounded in first order logic and includes constructs called semantic formulations that structure the meaning of rules.

The rest of the paper is structured as follows: In Section 2, we introduce the notion of business rules and CNLs. Section 3 presents some related work on CNLs for business rules and in Section 4, we describe the RuleCNL in detail. Section 5 presents the RuleCNL tool and Section 6 the conclusions and future work.

\section{Business Rules and Controlled Natural Languages}
\subsection{Business Rules}
In the literature, we find numerous definitions of business rules. However the most used definition is given by the Business Rule Group (BRG) \cite{BRGdef} as follows:

\textit{"a business rule is a statement that defines or constrains some aspect of the business.  It is intended to assert business structure, or to control or influence the behavior of the business."}

\textbf{E.g.} \textit{A loan must be approved if its value is less than 10,000 Euros.}

One challenge with the BRA is to find the characteristics of a good business rule. Some workers \cite{BRGapproach} \cite{Morgan} have proposed a set of  characteristics for a business rule statement to be deemed as good. Among them, we can cite that a business rule should be atomic, declarative, business related, consistent, unambiguous, etc.

In the context of the BRA \cite{Babara} \cite{Ronald2003}, many formal languages have been proposed by many vendors for business rules modeling. These languages have a well-defined syntax, an unambiguous semantics and support automated reasoning over rules. \cite{StateofArt}  provides a state of the art on business rule languages and concludes that most of them are hard to use for business people without training in formal methods, but are rather easy for software engineers. We contend that  business rules should be expressed declaratively in NL sentences for the business audience. Thus, CNLs are good solutions for bridging the gap between natural and formal languages \cite{Levy}.

\subsection{Controlled Natural Languages}

CNLs are engineered subsets of natural languages whose grammars and vocabularies have been restricted in a systematic way in order to reduce both the ambiguity and complexity of full NLs (e.g. English, French, etc.) \cite{Rolf2010}. 

In general, CNLs fall into two broad categories: human-oriented CNLs and machine-oriented CNLs.  Human-oriented CNLs are intended to improve the communication among people for specific purposes and the readability and comprehensibility of technical documentations.  They have no formal semantics and are usually defined by informal guidelines \cite{Kuhn2010}.
Machine-oriented CNLs are designed to improve the communication between humans and computers. They are completely unambiguous and can be defined by formal grammars with a direct mapping to formal logic \cite{Kuhn2010}. 

Machine-oriented CNLs can be used in various domains and applications such as CNL for knowledge representation (Attempto Controlled English \cite{ACE}, Processable English \cite{PENG}, Computer Processable Language \cite{CPL}, etc.), CNL for ontologies \cite{Rabbit}, CNL for semantic web \cite{Rolf2005}, CNL for machine translation \cite{CentreTesniere},  CNL for business rules like RuleCNL as  presented  in this paper, etc.

\section{Related Work}

The idea of verbalizing rules that already exist in a formal representation \cite{Halpin} \cite{Rewerse} has led the domain of business rules to become an interesting application of CNLs. Because business rules need to be approved and followed by people with no particular background in formal or logic representation, it is important to have an intuitive representation that CNLs can offer.  There are some CNLs that have been defined for this particular problem area.

In the business context, the Object Management Group (OMG)\footnote{http://www.omg.org/}  has published a  standard called SBVR \cite{SBVR}  which provides a means for describing the structure of the meaning of rules expressed in the natural language that business people use.  However, SBVR is not itself a CNL so it is up to each SBVR-implementing language or notation to specify its formal mechanisms \cite{ Anderson}. SBVR claims to be restricted to semantics leaving apart a key functionality of NLs, which is syntax. Thus, for various reasons, the SBVR standard did not include a normative specification of the language to be used by business people to express their vocabulary and rules. 

SBVR-Structured-English (SE) \cite{SBVR} and  RuleSpeak  \cite{RuleSpeak} are both defined in the SBVR specification as CNLs to express business rules in a restricted version of English. However, these CNLs are not languages per se, but rather a set of best practices for human  speakers. They are defined informally by sets of guidelines based on experiences of best practice in rule systems \cite{Kuhn2010}. They are not normative and have no formal grammar but can be mapped to the semantics formulation of the SBVR meta-model. They are not supported by any tooling and cannot be processed in a fully automatic way \cite{Ruth}. The syntax is achieved by text formatting and coloring, which could be used to aid  understanding by the domain expert user. However, a CNL requires a formal definition of its syntax (the language's grammar), which can be used to support business users in the process of entering syntactically correct inputs. This limitation is avoided by our RuleCNL controlled natural language.

\section{RuleCNL: A Controlled Natural Language for Business Rules Specifications}

\subsection{Introduction}

In this section, we present our RuleCNL for expressing business rules. As mentioned in the introduction, its methodology is based on the alignment of the business rule definitions with the business vocabulary. Thus, business rules are semantically connected to the business domain and readily understandable by  domain experts. The methodology is derived from the core idea of the BRA advocated by the BRG as follows: {\textit{"Rules build on facts, and facts build on concepts as expressed by terms."}\cite{BRGapproach}. In order to overcome the limitations highlighted in the previous section, we defined a formal grammar and therefore a parser that can be used for the syntax analysis of rules. The writing process of a business rule is fully supported by the consistency check imposed by the methodology. Its semantics is defined by automatic mappings into the SBVR semantic formulations.  This enables a language-independent way of describing the semantic structure of rule statements  and is grounded on a sound theoretical foundation of formal logic.  Fig. \ref{Fig. 1} shows the general architecture of the RuleCNL.

\begin{figure}[H]
\centering
\includegraphics[width=350pt, height = 150pt]{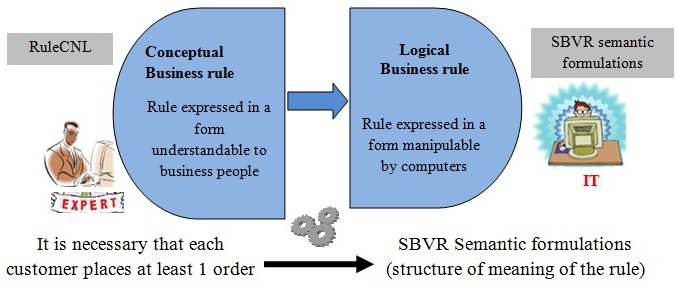}
\caption{General architecture of the RuleCNL}
\label{Fig. 1}
\end{figure}

 For the sake of readability, the SBVR semantic formulations of the example of the Fig. \ref{Fig. 1} is shown at the Fig. \ref{Fig. 2}

\begin{figure}[H]
\centering
\includegraphics[width=350pt, height = 200pt]{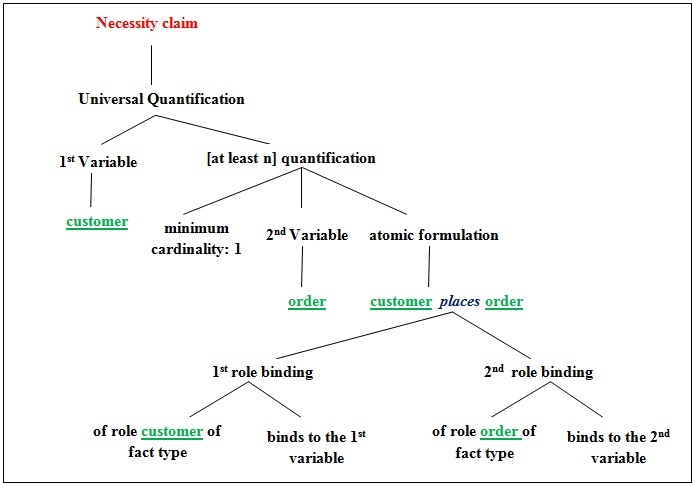}
\caption{SBVR semantic formulations of the rule shown in the Fig. 1}
\label{Fig. 2}
\end{figure}

\subsection{RuleCNL Vocabulary }

The RuleCNL vocabulary represents the conceptual model of the business domain which defines a cohesive set of interconnected concepts (domain terms and their relations) that a given company uses in its talking or writing in the course of doing business. It is defined in a structured way by a business user and represents the knowledge that the  company knows about itself. The RuleCNL is domain-independent and the vocabulary consists of:

\textbf{- \textit{Domain Term}:} designates a significant business entity  that can be represented by a common noun or a noun phrase. 
(e.g. \textit{customer}, \textit{gold customer}, \textit{bank account}, etc.). A \textit{Domain Term} is always represented in a singular form and with no articles or determiners.

\textbf{- \textit{Domain Name}:} designates a significant business entity that represents only one thing. It is usually a proper name. (e.g. \textit{ France}, \textit{Euro}, \textit{USA}, etc.)

\textbf{- \textit{Domain Verb}:} designates a relationship, situation, or action involving one or two  \textit{Domain Terms/Names}. In order to keep the RuleCNL vocabulary simple and readily manipulable by domain experts, we only consider unary and binary  \textit{Domain Verbs}. 

The binary \textit{Domain Verb} defines a semantic relationship and has two placeholders filled by \textit{Domain Terms/Names} and its declaration syntax is \textit{Subject + Domain Verb + Object}. The \textit{Domain Verb} per se is only a part of this declaration syntax and has no meaning in isolation, but only within the relationship.\\

For instance: Let us consider the verb \textit{to run}, it has different meanings within the following relationships:\\

\underline{manager} \textit{runs} \underline{company};
\underline{horse} \textit{runs} \underline{race};
\underline{computer} \textit{runs} \underline{program}\\

A \textit{Domain Verb}  can be written both in active or passive form. (e.g. \textit{customer places order}; \textit{order is placed by customer}). The \textit{Domain Verb} can be a linguistic verb (in this case, it is conjugated in the third-person  singular) or a combination of a verb with some functional words (preposition, etc.).

The unary \textit{Domain Verb} defines a characteristic or a state of a \textit{Domain Term/Name} and its declaration syntax is \textit{Subject + Domain Verb}. Its evaluation leads to a Boolean value. (e.g. \textit{order shipped}, \textit{customer smokes})

There are no additional words or functional words in the relationship. This leads to a great flexibility and any constraints or restrictions will be added when defining business rules. The RuleCNL vocabulary includes some built-in relationships as comparison verbs (equality/inequality) that are not defined by domain users. Domain users define or import their vocabulary with the help of the vocabulary editor.

\subsection{RuleCNL Grammar}

RuleCNL grammar defines syntax rules and constrains for business rules. It assumes the existence of RuleCNL vocabulary and makes reference to \textit{Domain Terms/Names} and \textit{Domain Verbs} defined in the vocabulary. The general structure of a rule is:  \textit{Modality + Statement}. 
The \textit{Modality} carries the sense of  operational or structural rules. (e.g. \textit{It is obligatory that}, \textit{It is necessary that}, etc.). 
The \textit{Statement} is a declarative sentence that regulates the structure of the rule. Its general pattern is a set of clauses which can go from a simple to a very complex/compound (linked by connectives and conditionals) structure. Each clause is always based on exactly one \textit{Domain Verb} defined in the vocabulary. Subsequently, it combines many linguistic particles (function words, adverbs, etc.) in order to form a  grammatically correct sentence. The grammar supports the use of complex noun phrases involving quantifications, instantiations and qualifications of \textit{Domain Terms}. It also supports verb phrases involving verbs and prepositional phrases and can be used to define simple, compound and conditional sentence structures. 
The following examples show some statement structures. In these examples, \textit{Domain Terms} are underlined; \textit{Domain Verbs} are in italic font and other linguistic particles are in bold font. 

\textbf{Simple statement:} It is based on only one clause. 

\textbf{E.g.} \textbf{each} \underline{customer}\textit{ places} \textbf{at least one} \underline{order} \\
 This statement follows the structure:  \textit{Subject + BinaryVerb + Object}  and is based on the clause (\textit{Domain Verb}) \underline{customer}\textit{ places} \underline{order}\\
\textit{Subject} and \textit{Object} are noun phrases with  determiners which in this case are quantifiers (each, at least). In general, a determiner can also map to an article (a, an, the) or nothing. \textit{Subject} and \textit{Object} make reference to \textit{Domain Terms} visible in the vocabulary. The \textit{BinaryVerb} also makes reference to a \textit{Domain Verb} of the vocabulary. In this example, \textit{Subject}, \textit{Object} and \textit{Domain Verb} are unqualified, and then the rule will be applied to any instance of the related \textit{Domain Terms}. However, they can be qualified by other descriptive elements, such as their existence in a particular state, in order to specify the applicability of the rule with enough precision. 

\textbf{Compound/complex statement:}  this is  recursively built from simpler statements through coordinators (and, or) and subordinators (if, who, that, which).

\textbf{E.g.} \textbf{each} \underline{order} \textit{is shipped} \textbf{ if the}  \underline{customer} \textbf{who} \textit{places} \textbf{the}  \underline{order} \textit{is adult} \textbf{and} \textit{holds} \textbf{an}  \underline{account} \textbf{that} \textit{has} \textbf{a}  \underline{outstanding balance} \textbf{that} \textit{is greater than} \textbf{0}\\
This statement is based on the following  \textit{Domain Verbs}:\\ 
\underline{order}\textit{ shipped};\\
 \underline{customer}\textit{ adult}; \\
 \underline{customer}\textit{ places} \underline{order}; \\ 
\underline{customer}\textit{ holds} \underline{account};  \\ 
\underline{account}\textit{ has} \underline{ outstanding balance} \\defined in the vocabulary \\
and the built-in comparison verb  \underline{quantity1}\textit{ is greater than} \underline{quantity2}

\subsection{Formalization of the Grammar}

In order to build a parser for the RuleCNL, we have formalized our grammar rules using Extended Backus-Naur Form (EBNF) which is a notation for specifying context free grammars. 
The RuleCNL grammar consists of a set of rewriting rules used to restrict the syntax of rule statements.
An excerpt of the general rule pattern is shown at Fig. \ref{Fig. 3}. The vertical bar (\textbar) is the disjunction and the comma character ( , ) is the conjunction.  The symbol ( )+ means that at least one occurrence of rule element enclosed in the bracket must appear at that point.

\begin{figure}[H]
\centering
\includegraphics[width=300pt,  height = 160pt]{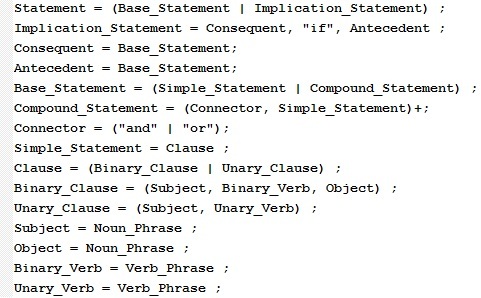}
\caption{General rule pattern}
\label{Fig. 3}
\end{figure}

\subsection{RuleCNL Semantics}

The RuleCNL semantics is defined by automatic mappings from RuleCNL rules to the SBVR semantic formulation (SF) description. SF is a part of the SBVR specification \cite{SBVR} that provides a means for describing the structure of the meaning of rules expressed in NL that business people use. SF is an abstract and language independent syntax to represent the meaning of a rule in a set of logic structures so that it can be machine processed. The full SBVR SF is not presented in this paper, but can be found in \cite{SBVR}.

 Fig. \ref{Fig. 4} shows the resulting SF in the XML form generated  by the RuleCNL tool from the example of Rule 1  below

Rule 1: It is obligatory that the customer "John" places at least one order 

\begin{figure}[H]
\centering
\includegraphics[width=360pt, height = 150pt]{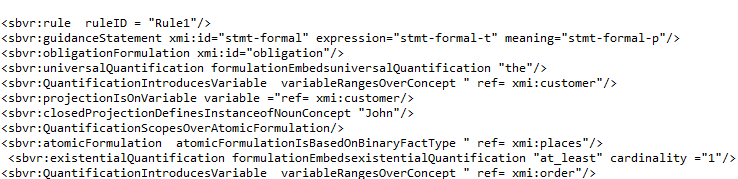}
\caption{Mapping of a RuleCNL rule to SBVR semantic formulation}
\label{Fig. 4}
\end{figure}

\textit{Domain Terms} are mapped to Noun concepts, determiners are mapped to quantifications, \textit{Domain Names} are mapped to Individual concepts, \textit{Domain Verbs} are mapped to Fact types, coordinators are mapped to logical operators, relative clauses are mapped to Projections, etc.

\section{RuleCNL Tool}
\subsection{Implementation}

An important feature of a reliable CNL is its tool support because one of the biggest problems (if not the biggest problem) of CNLs is the usability of a new CNL by end-users  \cite{Kuhn2010}. In reality, the limited expressiveness due to the restriction on vocabulary and grammar of CNLs  leads to the difficulty in writing statements that comply with the imposed restriction. Writing syntactically correct statements without tool support is much more complicated because the user needs to learn syntax restrictions, which are in many cases not trivial to explain. Thus, we have developed  editors that make the writing process and the usability of the RuleCNL as effortless as possible. There are two editors: a vocabulary editor (Fig. \ref{Fig. 5}) and a rule editor (Fig. \ref{Fig. 6})

\begin{figure}[H]
\centering
\includegraphics[width=360pt, height = 200pt]{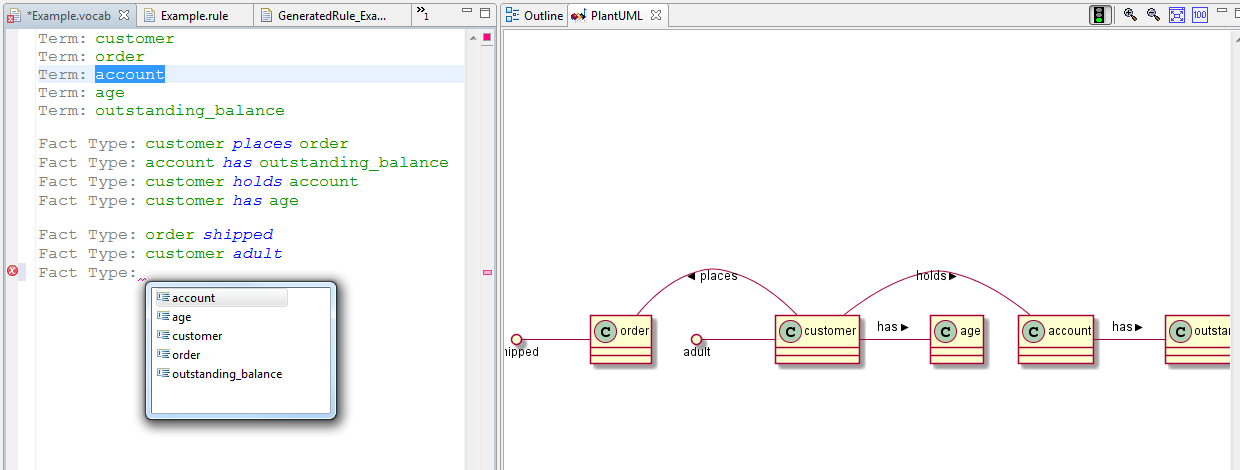}
\caption{Vocabulary editor}
\label{Fig. 5}
\end{figure}

\begin{figure}[H]
\centering
\includegraphics[width=360pt, height = 250pt]{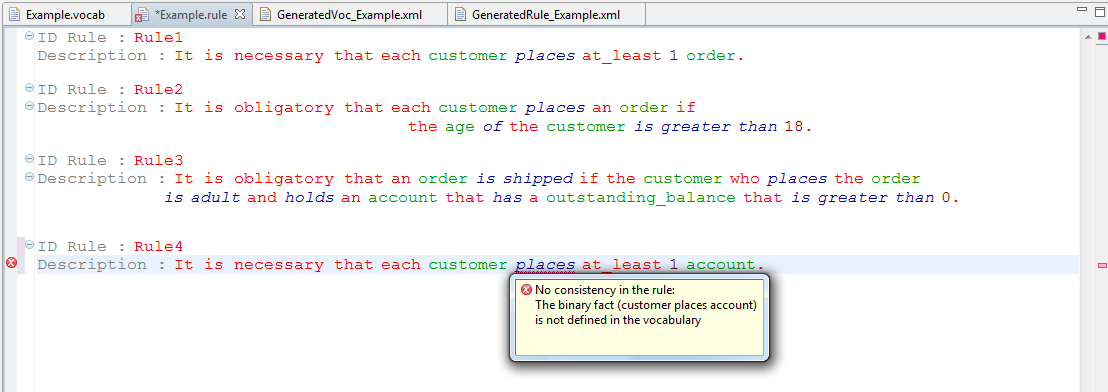}
\caption{Rule editor}
\label{Fig. 6}
\end{figure}

The vocabulary editor assists the users to specify their vocabulary. It has two customized and dynamic views on the vocabulary in use:  an outline view (tree-like) and a graphical view (UML-like notation). The rule editor assists the users to specify the business rule statements.

Both editors offer high level features such as auto-completion, error-handling, automatic highlighting and validation.
\subsection{Evaluation}

An  ideal  CNL  should  be  effortless  to  learn and  expressive  enough  to  describe the domain  problem.  Thus, we have evaluated the RuleCNL with respect to its expressivity  and comprehensibility. For the experiment, we collected about 50 business rules from real-life case studies of two companies written in the English and  French languages. The first company operates in the domain of banking and insurance whereas the second is a parastatal.  The evaluation was carried out by four end-users divided into two groups: group 1 is  made up of two business experts with no background in formal notations of business rules and  group 2 is  made up of two business users with a background in information system technology. Our  objective  was twofold and  consisted  in  finding how many  business  rules  the RuleCNL  could formalize in a natural way and  how easy the  users  could understand the formalization. Thus, the evaluation's metrics are the expressiveness and the users’ comprehension  of the RuleCNL. One could have also added the readability, but as  RuleCNL is close to  NL, statements written in RuleCNL are read in the same way as in the underlying NL.   Table \ref{Table 1} shows the result of the experiment with the agreement of the two users of each group. This result is the same both for English and French.

\begin{table}[H]
\begin{center}
\caption{Evaluation }
\label{Table 1}
\begin{tabular}{|p{3cm}|p{3cm}|p{3cm}|}
\hline
 \textbf{Measures/Users} &  \textbf{Group 1}  & \textbf{Group 2} \\
 \hline \textit{Expressiveness} & 84\%  & 84\%   \\
 \hline \textit{Comprehensibility}	& 90\%	& 100\% \\
 \hline
\end{tabular}
\end{center}
\end{table}

As we can see in the Table 1, the expressiveness is 84\% for both groups. The remaining 16\% was because of  syntactic and semantic ambiguities in some rules. However, with more training, the users rephrased these rules so that the tool was able to formalize.  Regarding the comprehensibility, group 1 confirmed that it understood 90\% of the formalized rules in a natural way and  group 2 understood all the rules. This result is not surprising because  group 1 users do not have a background in formal notations. Thus, the 10\% remaining consisted of complex rules, which require much constrains imposed by the grammar. 

\section {Conclusions and Future Work}

The ultimate goal to bridge the gap between natural and formal languages has brought interesting research challenges in the area of CNL. In this paper, we have presented the RuleCNL which is a CNL for business rule specifications.  The aim of RuleCNL  is to help business experts  formalize their business rules in a business-friendly way that can be understood by computers. The RuleCNL syntax is defined by a formal grammar and its semantics is defined by automatic mappings of RuleCNL rules to SBVR semantic formulations. The RuleCNL tool provides  editors that assist end-users in the writing process. RuleCNL along with its tool have been evaluated with satisfactory results from business experts. We are currently improving the tool and extending evaluation to many other companies. The future work will be to go from the SBVR SF to some production rules for rule engines or software components.

\end{document}